\documentclass{IEEEcsmag}

\usepackage[colorlinks,urlcolor=blue,linkcolor=blue,citecolor=blue]{hyperref}
\expandafter\def\expandafter\UrlBreaks\expandafter{\UrlBreaks\do\/\do\*\do\-\do\~\do\'\do\"\do\-}
\usepackage{upmath,color}

\usepackage{threeparttable} 
\usepackage{makecell}
\usepackage{multirow}
\usepackage{listings}
\usepackage{tablefootnote}
\usepackage[superscript]{cite}
\usepackage[gen]{eurosym}

\definecolor{backcolour}{rgb}{0.95,0.95,0.92}
\definecolor{green}{rgb}{0,0.6,0}
\definecolor{gray}{rgb}{0.5,0.5,0.5}
\definecolor{mauve}{rgb}{0.58,0,0.82}
\definecolor{darkgray}{rgb}{.4,.4,.4}

\lstdefinelanguage{JavaScript}{
  keywords={typeof, new, true, false, catch, function, return, null, catch, switch, var, if, in, while, do, else, case, break},
  keywordstyle=\color{blue}\bfseries,
  ndkeywords={class, export, boolean, throw, implements, import, this},
  ndkeywordstyle=\color{darkgray}\bfseries,
  identifierstyle=\color{black},
  sensitive=false,
  comment=[l]{//},
  morecomment=[s]{/*}{*/},
  commentstyle=\color{purple}\ttfamily,
  stringstyle=\color{red}\ttfamily,
  morestring=[b]',
  morestring=[b]"
}

\lstdefinestyle{code}{
    language=JavaScript,
    basicstyle=\ttfamily\small,
    commentstyle=\color{green},
    keywordstyle=\color{blue},
    numberstyle=\tiny\color{gray},
    stringstyle=\color{mauve},
    breakatwhitespace=false,
    breaklines=true,
    captionpos=b,
    keepspaces=true,
    numbers=left,
    numbersep=5pt,
    showspaces=false,
    showstringspaces=false,
    showtabs=false,
    tabsize=2,
    escapechar=\#,
}

\jvol{XX}
\jnum{XX}
\paper{8}
\jmonth{Month}
\jname{Publication Name}
\jtitle{Publication Title}
\pubyear{2024}

\AtBeginDocument{%
  }

\begin{document}


\sptitle{Theme Article: Special Issue on Next-generation Software Testing: AI-powered Test Automation}


\title{APITestGenie: Automated API
Test Generation through Generative AI}


\author{André Pereira}
\affil{FEUP, Porto, Portugal}

\author{Bruno Lima}
\affil{FEUP, UMaia and LIACC, Porto, Portugal}

\author{João Pascoal Faria}
\affil{FEUP and INESC TEC, Porto, Portugal}

\markboth{THEME/FEATURE/DEPARTMENT}{THEME/FEATURE/DEPARTMENT}

\begin{abstract} 
Intelligent assistants powered by Large Language Models (LLMs) can generate program and test code with high accuracy, boosting developers' and testers' productivity.
However, there is a lack of studies exploring LLMs for testing Web APIs, which constitute fundamental building blocks of modern software systems and pose significant test challenges. 
Hence, in this article, we introduce \textit{APITestGenie}, an approach and tool that leverages LLMs to generate executable API test scripts from business requirements and API specifications.
In experiments with 10 real-world APIs, the tool generated valid test scripts 57\% of the time. With three generation attempts per task, this success rate increased to 80\%.
Human intervention is recommended to validate or refine generated scripts before integration into CI/CD pipelines, positioning our tool as a productivity assistant rather than a replacement for testers. Feedback from industry specialists indicated a strong interest in adopting our tool for improving the API test process.

\end{abstract}

\maketitle

\chapteri{G}enerative Artificial Intelligence (AI) has witnessed remarkable progress in recent years, reshaping the landscape of intelligent systems with applications in many domains. Studies show that intelligent assistants powered by Large Language Models (LLMs) can now generate program and test code with high accuracy based on textual prompts~\cite{ref:copilotSimpleProblems}, boosting developers' and testers' productivity.


Although LLMs have been explored for several test generation tasks, there is a lack of studies exploring them for testing Web APIs, which constitute fundamental building blocks of modern software systems.  
Guaranteeing the proper functioning of API services has become substantially more challenging due to the significant increase in the number of APIs in recent years~\cite{ref:stateAPI2023}.

Identifying and creating relevant API tests is a tedious, time-consuming task 
\cite{ref:testAutomation}, 
requiring significant programming expertise, particularly as system complexity increases. While integration and system tests ensure functional integrity, they are more challenging to generate due to their abstract nature. Existing API test automation tools often fail to leverage high-level requirements to produce comprehensive test cases and detect semantic faults.


To overcome such limitations by taking advantage of the recent developments in Generative AI, we present APITestGenie -- a tool that leverages LLMs to generate executable API test scripts from business requirements written in natural language and API specifications documented using OpenAPI\cite{ref:openAPIonline} the most widely adopted standard for documenting API services.

The main research questions we aim to address in the development and evaluation of APITestGenie are:
\vspace*{-6pt}
\begin{itemize}
    \item[{\ieeeguilsinglright}] \textbf{RQ1}---How effectively and efficiently can LLMs be used to generate API test scripts from business requirements and API specifications?
    \item[{\ieeeguilsinglright}] \textbf{RQ2}---Can the industry adopt Generative AI to assist in API test generation?
\end{itemize}
\vspace*{-5pt}

In the next sections, we analyze related work, present our solution and its evaluation, and point out the main conclusions and future work.

\vspace*{-10pt}

\section{RELATED WORK}


In recent years, several approaches and tools have emerged for testing REST APIs, with a significant rise in research since 2017, focusing on test automation using black-box techniques and OpenAPI schemas~\cite{golmohammadi2023testing}. However, these approaches, including AI-driven ones, have not utilized LLMs.

In a 2022 study \cite{kim2022automated}, the authors compare the performance of 10 state-of-the-art REST API testing tools from industry and academia on a benchmark of 20 real-world open-source APIs, in terms of code coverage and failures triggered (5xx status codes), signaling out EvoMaster among the best-performent.


EvoMaster\cite{ref:evoMasterOnline} can generate test cases for REST APIs in both white-box and black-box testing mode, with worse results in the former due to the lack of code analysis. It works by evolving test cases from an initial population of random ones, trying to maximize measures like code coverage (only in white-box mode) and fault detection, using several kinds of AI heuristics~\cite{ref:apiAutomatedTestGeneration}. Potential faults considered for fault finding are based on HTTP responses with 5xx status codes and discrepancies between the API responses and what is expected based on the OpenAPI schemas.


However, like other approaches that rely solely on the OpenAPI specs as input \cite{ref:quickRest,ref:gpt_sbst, ref:apiBlackBoxTesting}, EvoMaster is limited by the absence of requirements specifications and context awareness, which restricts the variety of tests that can be generated and the types of faults that can be detected. Our goal is to leverage requirements specifications and the context awareness capabilities of LLMs to overcome these limitations, enabling the generation of more comprehensive and effective tests.

Recent work has started leveraging LLMs for REST API testing. RESTGPT~\cite{kim2024leveraging} enhances testing by extracting rules from OpenAPI descriptions but does not generate executable scripts. RESTSpecIT~\cite{decrop2024you} uses LLMs for automated specification inference, discovering undocumented routes and query parameters, and identifying server errors.
Both approaches complement ours, as the enhanced or inferred specifications can be used as input for subsequent test script generation.  

To our knowledge, APITestGenie is the first approach that leverages the context awareness and natural language processing capabilities of LLMs to generate executable scripts for testing REST APIs, based on the specification of the business requirements in natural language and the specification of the APIs under test in OpenAPI, overcome some limitations of prior methods.

\section{SOLUTION}





\subsection{Architecture and Workflow}

\begin{figure}
\centerline{\includegraphics[width=18.5pc]{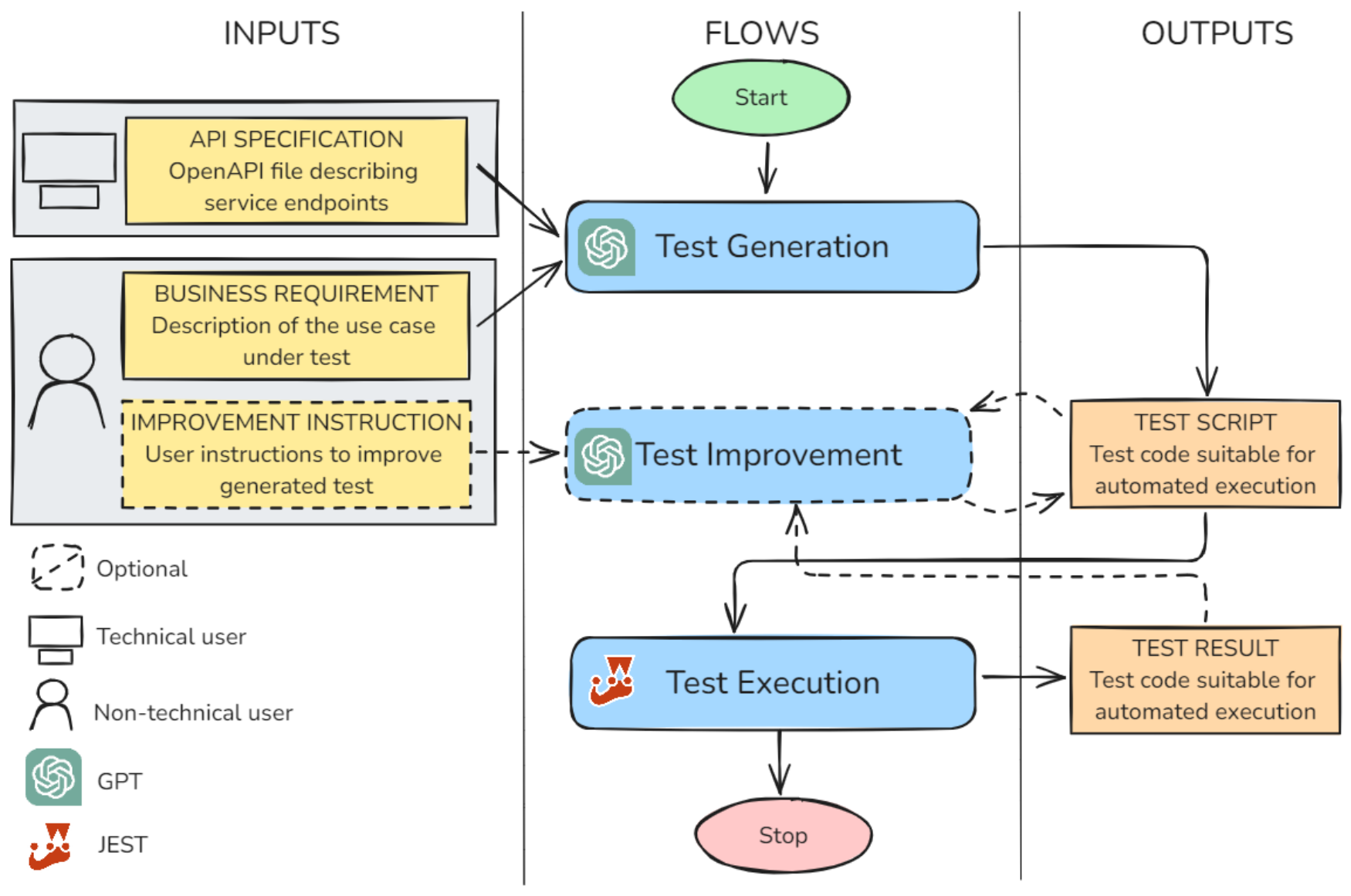}}
\caption{APITestGenie flow diagram, showcasing the interactions of the main flows in the system, inputs and outputs.}
\label{fig:userArchitecture}
\vspace*{-10pt}
\end{figure}

APITestGenie autonomously generates, improves, and executes test cases. Its architecture (see Figure~\ref{fig:userArchitecture}) is divided into three modular flows: Test Generation, Test Improvement, and Test Execution. While each flow can operate independently, they are designed to work together for a more complete solution.

The Test Generation flow receives as input the API Specification and the business requirement to generate a test script. 
Depending on the API size, we use either retrieval augmented generation (RAG) for larger API specs or full processing for smaller ones. We then construct system and user prompts based on the TELeR taxonomy~\cite{ref:teler} for a <single turn, instruction, high detail, defined> use case.
Lastly, we call the LLM and parse the content generated, outputting the test script. An example of a generated test script can be found in our GitHub at \url{https://github.com/Andrepereira2001/APITestGenie_CatFactTestScript/blob/main/test.ts}. 

The Test Improvement flow can be used to improve the generated script, based on previous test results and improvement user instructions. 
The Test Execution flow uses JEST
in a Typescript environment to execute the generated test script and report the results.



\vspace*{-10pt}
\subsection{System and User Prompt}

The system prompt is structured as follows:
\vspace*{-5pt}
\begin{enumerate}
    \item {\it Context}---specifies the task to be performed.
    \vspace*{-5pt}
    \begin{enumerate}
        \item High-level goal.
        \item Target user and objective guidelines.
        \item Test structure example
        \item Information on available environment variables.
    \end{enumerate}
    \item {\it Performance}---guidelines describing how the LLM will be evaluated.
    \vspace*{-5pt}
    \begin{enumerate}
        \item Output evaluation metrics.
        \item Generation guidelines.
    \end{enumerate}
    \item {\it Output}---details how the model output should be formulated.
    \vspace*{-5pt}
    \begin{enumerate}
        \item Reasoning steps.
        \item Output components and structure.
    \end{enumerate}
\end{enumerate}
The user prompt is structured as follows:
\vspace*{-5pt}
\begin{enumerate}
    \item {\it Requirement}---business requirement under test.
    \item {\it API}---API specification under test.
\end{enumerate}\vspace*{-5pt}


\vspace*{-10pt}
\subsection{Prompt with the API Specification}

A significant limitation of current LLM models is their context window size. 
Therefore, we created a script that simplifies the raw API specification by removing all the \textit{<img>} tags content and all the \textit{admin} and \textit{deprecated} resources.

Additionally, we use RAG~\cite{ref:metaRAG} to filter relevant information, allowing us to test APIs of any size. To do so, we first slice the API specifications into chunks according to JSON syntax. Then, we expand the business requirement into five versions used in the RAG process to retrieve a set of similar chunks.

\vspace*{-10pt}
\subsection{Prompt with the Business Requirement}





A key input for test generation, besides the API specification, is the prompt provided by the user (or fetched from a tool) with the description of the business requirement to be tested. 



\begin{figure}
\centerline{\includegraphics[width=18.5pc]{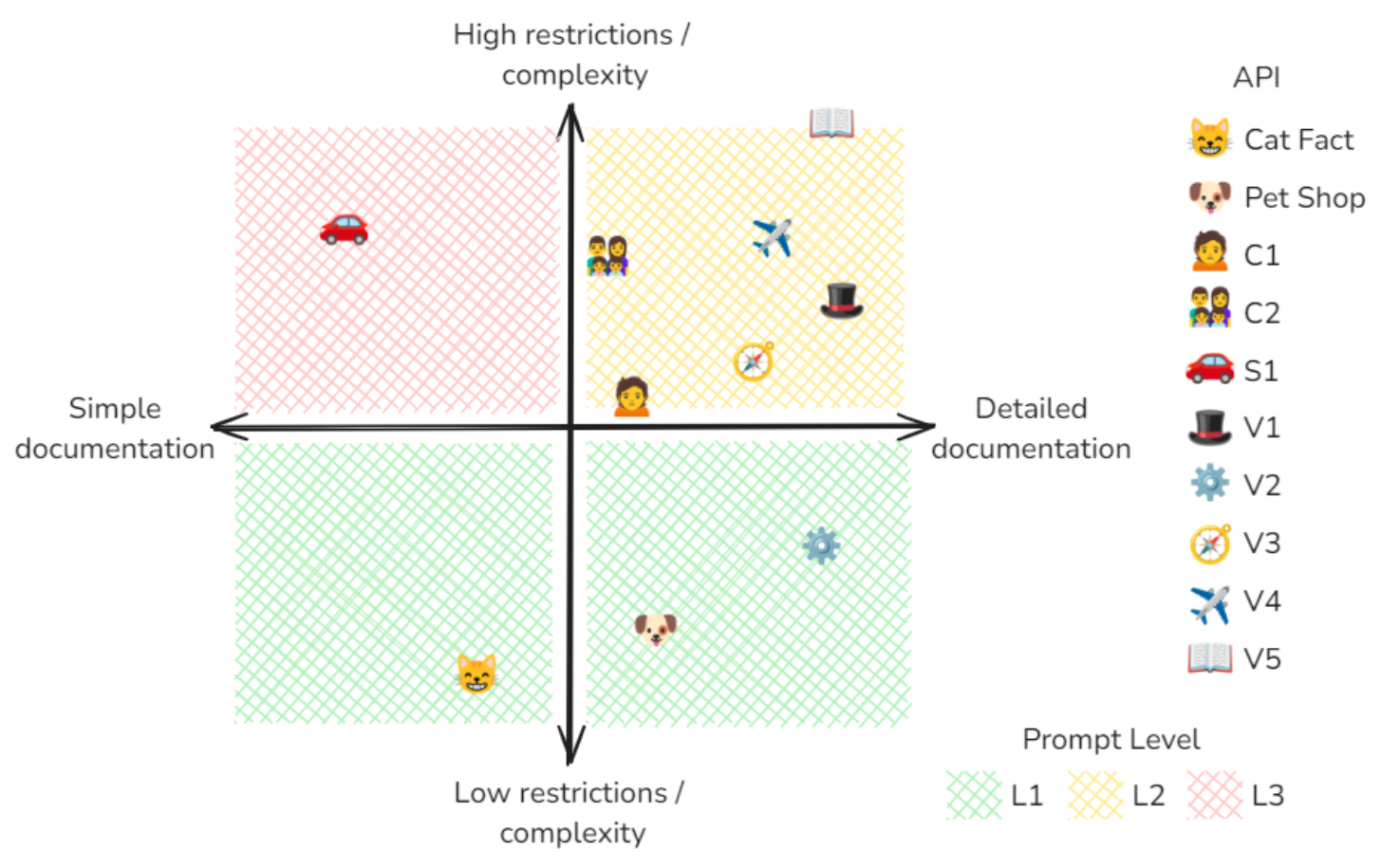}}
\caption{Characterization of the APIs under test based on their internal complexity and the level of detail of the documentation, with associated prompt levels recommended.} 
\label{fig:apiMatrix}
\vspace*{-10pt}
\end{figure}

Experiments showed that detailed business requirements are crucial to test complex or poorly documented APIs. We classify APIs along two dimensions (Figure~\ref{fig:apiMatrix}): documentation detail and complexity/restrictions. The horizontal axis assesses the thoroughness of resource explanations, while the vertical axis evaluates API complexity and security constraints. Each quadrant suggests a corresponding prompt level.

Three prompt levels are recommended as follows:

\begin{enumerate}
    \item {\it LEVEL 1}---Basic requirement description
    \begin{itemize}
        \item[{\ieeeguilsinglright}]Basic description of a business requirement, typically using a user story structure. 
        \item[{\ieeeguilsinglright}] {\it Example}---As a user, I want to receive a new and random cat fact every time I open the app or refresh the content so that I can learn intriguing information about cats and stay engaged with the app.
    \end{itemize}
    \item {\it LEVEL 2}---Experimental data inclusion
    \begin{itemize}
        \item[{\ieeeguilsinglright}]Adds experimental data to use in the API request, such as parameter values, or relations between inputs and expected outputs.
        \item[{\ieeeguilsinglright}] {\it Example}---As a user, I want to retrieve all the available models in the eu27+ country so that I can know which vehicles are available.
    \end{itemize}
    \item {\it LEVEL 3}---Comprehensive guiding information
    \begin{itemize}
        \item[{\ieeeguilsinglright}]Adds guidance, such as test plans, acceptance criteria, or tips to use API resources.
        \item[{\ieeeguilsinglright}] {\it Example}--- As a user, I want to retrieve the price of accessories for a car in the Italian market. First, use the vehicle's service to get a list of all available vehicle models in the market. Then, using the accessories service and with a random model previously selected, retrieve the list of all available accessories on 17/05/2024. Lastly, use the pricing service to calculate the price of the available car accessories.
    \end{itemize}
\end{enumerate}\vspace*{-10pt}

\vspace*{-10pt}

\section{EVALUATION}

\begin{table*}[h!]
    \centering
    \begin{tabular}{l|c|c|c|c|c|c|c}
        \hline
        \textbf{Service$^a$} & \textbf{Access} & \textbf{Original} & \textbf{Simplified} & \textbf{User Prompts} & \textbf{Generated Test} & \textbf{Valid Test} & \textbf{Generation}\\
        &  & \textbf{tokens$^b$} & \textbf{tokens$^c$} & \textbf{\texttt{(L1,L2,L3)}$^d$}& \textbf{Scripts$^e$} & \textbf{Scripts} & \textbf{Time (avg.)} \\ \hline
        Cat Fact$^f$ & Public & 754 & 754 & 1 \texttt{(1,-,-)} & 3 & 2 & 92s \\
        Pet Shop$^g$ &  Public & 4,070 & 4,070 & 2 \texttt{(2,-,-)} & 6 & 4 & 75s \\
        C1 &  Private & 25,966 & 25,475 & 2 \texttt{(1,1,-)} & 6 & 5 & 113s \\
        C2 & Private & 47,497 & 47,004 &  2 \texttt{(-,2,-)} & 6 & 3 & 98s \\
        S1 & Private & 49,761 & 49,761 & 2 \texttt{(-,-,2)} & 6 & 0 & 154s \\
        V1 & Private & 882,272 & 14,018 & 2 \texttt{(2,-,-)} & 6 & 3 & 115s \\
        V2 & Private & 430,739 & 16,094 & 5 \texttt{(2,1,2)} & 15 & 15 & 102s \\
        V3 & Private & 1,620,488 & 82,267 & 2 \texttt{(1,1,-)} & 6 & 2 & 131s \\
        V4 & Private & 417,109 & 108,724$^h$ & 4 \texttt{(2,1,1)} & 12 & 8 & 146s \\
        V5 & Private &  5,021,555 & 424,465$^h$ & 3 \texttt{(-,-,3)} & 9 & 1 & 227s \\ 
        \hline
        \textbf{Total} & & \textbf{9,254,211} & \textbf{772,632} & \textbf{25 \texttt{(11,6,8)}} &  \textbf{75} & \textbf{43 (57.3\%)} & \textbf{126s} \\
        \hline
    \end{tabular}
    \begin{tablenotes}
    \footnotesize
    \item a. Private API names from our automotive partner are just identified by the initials C-customer, S-shopping, and V-vehicle.
    \item b. Number of tokens in the original OpenAPI specification of the service under test counted with the tiktoken tokenizer.
    \item c. Number of tokens in the simplified spec (used as input for test generation), after removing irrelevant tags and resources. 
    \item d. Number of business requirements tested, discriminating the level of detail (L1 to L3) of the user prompts elaborated.
    \item e. Three test script generation attempts were performed for each user prompt.
    \item f. Available at \url{https://catfact.ninja/}.
    \item g. Available at \url{https://petstore.swagger.io/}.
    \item h. RAG was used to filter simplified API specs with over 100K tokens, reducing them to fit the LLM's maximum prompt size.
    \end{tablenotes}
    \vspace{1pt}
    \caption{Experimental API test generation results. In 75 generation attempts for 10 different APIs, 57.3\% of the generated test scripts were valid (further details in Table \ref{tab:generationAssessment}). The average generation time was 126s per script.}
    \label{tab:generationByService}
    \vspace*{-5pt}

\end{table*}

We evaluated APITestGenie in two phases. In the first phase, we evaluated its performance by collecting metrics to assess the correctness of the generated test scripts and the generation costs. In the second phase, we conducted a practical workshop with our industry partner to gather user feedback on the tool's capabilities and industry adoption. 

\vspace*{-10pt}

\subsection{Experimental Evaluation}

We assessed the robustness of test scripts generated by APITestGenie for 10 APIs with varying complexity and documentation levels (see Figure~\ref{fig:apiMatrix} and Table~\ref{tab:generationByService}). We used 25 unique user prompts as input, repeating the generation process three times per prompt, resulting in 75 test scripts.
Using GPT-4-Turbo with a 128k token Window, the average time and cost per test generation were 126s and \euro{0.37}, respectively.

We then executed the generated test scripts, identifying seven distinct result types, each with specific errors and resolution methods (see Table~\ref{tab:generationAssessment}). APITestGenie successfully generated valid test scripts 57.3\% of the time, with 12.0\% invalid due to external factors and 30.7\% due to generator limitations. 
The 75 scripts contained 90 test cases, 68\% of which were valid.


\begin{table*}[h!]
    \centering
    \begin{tabular}{p{1.8cm}|c|p{6cm}|p{4cm}}
        \hline
        \hline
        \textbf{Result type} & \textbf{Test Scripts} & \textbf{Description} & \textbf{Resolution} \\ 
        \hline
                \hline

        Pass \& valid $^a$ & 31 & Test execution succeeds, and the script is considered valid in manual inspection. & Save test script for regression testing. \\
\hline
        API defect & 12 & Test execution fails due to a defect in the API specification or implementation. The failures identified include inconsistencies with requirements in test cases involving multiple endpoints. & Fix API specification or implementation and re-run test script. \\
\hline
\hline
        Environment setup error & 7 & Test execution fails due to contamination of the generated script by incorrect environment setup (e.g., access tokens). & Fix environment setup and re-generate test script. \\      
\hline
        Missing API information & 2 & Test execution fails because key information needed for generating valid tests is missing in the API specification. & Fix API specification and re-generate test script. \\ 
        \hline
        \hline
        Hallucination & 19 & Test execution fails because of LLM hallucinations: wrong or non-existent import,  attribute, data manipulation operation, response structure validation, etc. & Fix test script, re-generate test script (possibly with a refined user prompt), or run test improvement flow. \\
        \hline
        Syntax error & 3 & The generated test script cannot be executed due to syntax errors. & Same as above. \\
\hline
        Empty script & 1 & No test script is generated. & Re-generate test script $^b$. \\

        \hline
        \hline
        \textbf{Total} & 75 & & \\
        \hline
        \hline
    \end{tabular}
    \begin{tablenotes}
    \footnotesize
    \item a. In our experiment, all the test scripts that executed successfully and passed were considered valid in the manual inspection.
    \item b. In our experiment, this was sufficient to generate a non-empty test script.
    \end{tablenotes}
    \vspace{1pt}
    \caption{Assessment of the generated test scripts. The first two cases represent valid test scripts (57.3\%). The remaining ones represent invalid test scripts, either due to factors external to the generator (12.0\%) or limitations of the generator (30.7\%). }
    \label{tab:generationAssessment}
    \vspace*{-5pt}
\end{table*}

\begin{figure}
\centerline{\includegraphics[width=18.5pc]{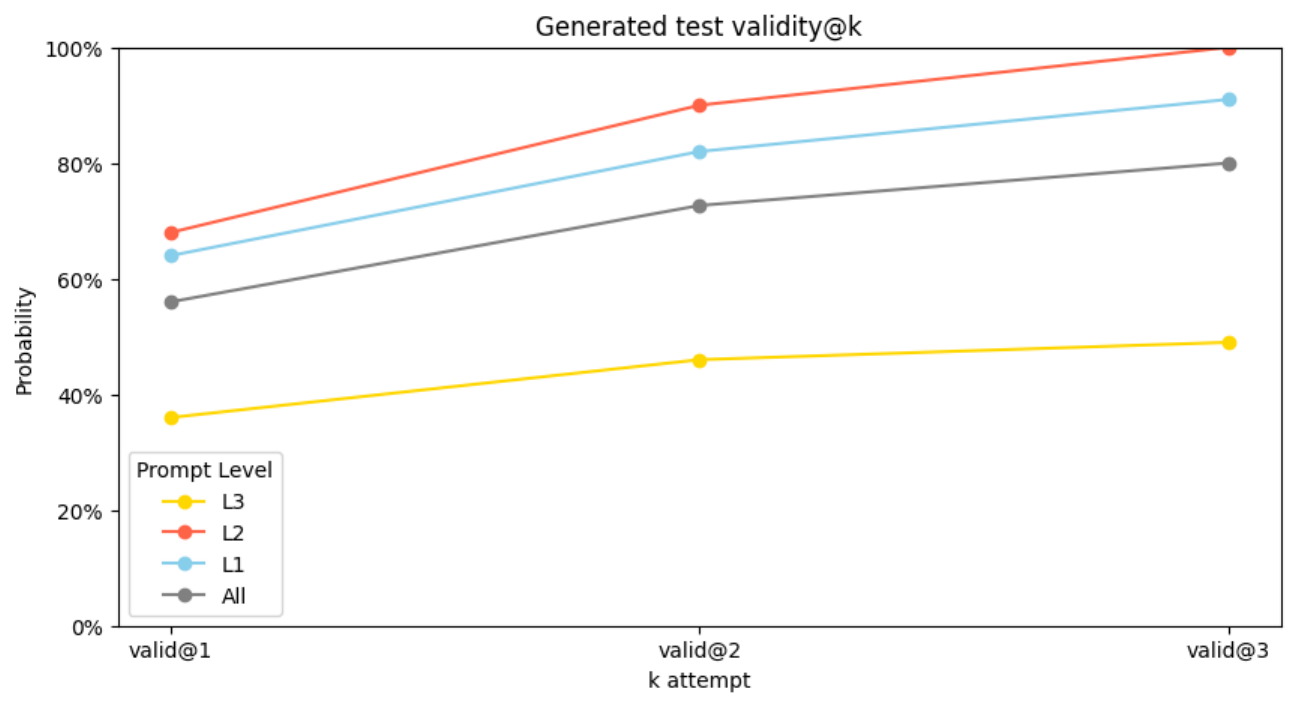}}
\caption{Estimated test generation success probability by number of attempts and prompt level (L1 to L3) or overall (All). $valid@k$ shows the average probability of generating at least one valid test script in $k$ attempts across all test generation tasks. Overall, $valid@1 \approx 57.3\%$ and $valid@3 \approx 80\%$.}

\label{fig:validk}
\vspace*{-10pt}
\end{figure}

Figure~\ref{fig:validk} shows that the probability of generating a valid test script increases from 57.3\% with one attempt to 80\% with three attempts. L2 prompts perform better than L1, likely due to the inclusion of test data. L3 prompts, used for more complex APIs, show less improvement. 

We found that API complexity affects test case generation results, but good documentation and tailored business requirements can improve tool performance. Most issues were related to semantic errors, not syntax. Despite this, the test structure remains useful, providing a solid starting point. 

\vspace*{-10pt}

\subsection{User Feedback}

\begin{figure}
\centerline{\includegraphics[width=18.5pc]{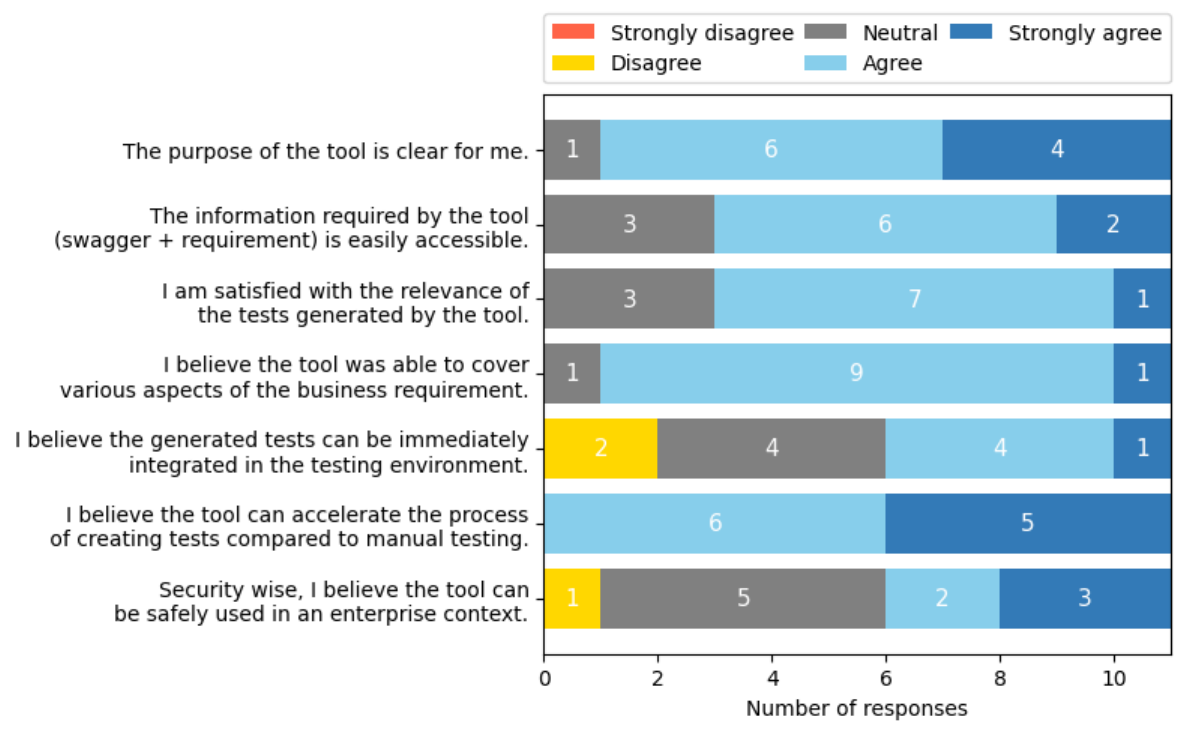}}
\caption{Staff opinions on APITestGenie generated tests. Opinions from 11 industry experts from our industry partner with over five years of experience in the product were collected anonymously after a workshop presentation.}
\label{fig:believes}
\vspace*{-20pt}
\end{figure}

We conducted a hands-on APITestGenie workshop with technical staff from our industry partner, who provided informed feedback on the relevance and completeness of the generated tests. During the workshop, we created four unique test scripts, demonstrating the tool's ability to handle progressively more complex requirements and APIs. Out of 15 participants, 11 responded to an anonymous survey (see Figure~\ref{fig:believes}).





Overall, APITestGenie was well received, with most collaborators finding the tests complete and relevant and showing interest in using the tool daily. However, some participants noted the need for improvements in data security and integration with existing testing environments.
\vspace*{-5pt}

\subsection{Answers to Research Questions}

Based on the evaluation conducted, we can answer the research questions as follows.

\textbf{RQ1:} LLMs can effectively be used to generate executable API test scripts from business requirements and API specifications, achieving up to 80\% test validity with clear prompts in three attempts. The process is efficient (126 seconds per test case), requiring minimal human intervention, though performance depends on high-quality business requirements and API specifications, with increased complexity requiring more human involvement.


\textbf{RQ2:} Generative AI tools like APITestGenie can enhance test generation efficiency by automating complex tasks, but adoption faces challenges due to security concerns, given the reliance on external closed-source LLMs, and the need for tight integration into existing workflows.




These conclusions are affected by the following threats to validity:
\vspace*{-5pt}
\begin{itemize}
    \item[{\ieeeguilsinglright}] Results may not generalize well to other domains or industries.
    \item[{\ieeeguilsinglright}] The evaluation metrics may not fully capture the quality of the generated test scripts.
    \item[{\ieeeguilsinglright}] Feedback from the workshop is statistically limited due to the small sample size.
\end{itemize}




\section{CONCLUSION}

We presented \textit{APITestGenie}, an approach and tool that leverages LLMs and RAG to generate executable test scripts from business requirements and API specifications for API testing.

In experiments with 10 real-world APIs, the tool generated valid test scripts 57.3\% of the time. With three generation attempts per task, this success rate increased to 80\%.
Feedback from an automotive industry partner indicated a strong interest in adopting our tool to improve their testing process.

Human intervention is recommended to validate or refine generated scripts before integration into CI/CD pipelines, positioning our tool as a productivity assistant rather than a replacement for testers.

Our results are not directly comparable to other tools in related work because of the different inputs used and our focus on generating tests that validate system requirements. 
Unlike traditional tools that typically focus on single endpoint testing, our approach leverages business requirements to generate tests that simulate real-world scenarios, often involving multiple endpoints, enabling comprehensive integration tests that ensure accurate service interactions and data flow. This makes our approach complementary to others.


Future work includes tighter integration into industrial workflows, collecting further metrics, applying multi-agent systems, and exploring open-source models to address cost and privacy concerns.


\vspace*{-10pt}



\def\refname{REFERENCES}


\bibliographystyle{ieeetr}
\bibliography{myrefs.bib}

\vspace*{-10pt}

\begin{IEEEbiography}{André Pereira} {\,} is a graduated master's student in Informatics and Computing Engineering at the Faculty of Engineering of the University of Porto (FEUP). His current research interests include LLM, Generative AI, and agent systems. Contact him at adbp@live.com.pt.
\end{IEEEbiography}

\vspace*{-10pt}

\begin{IEEEbiography}{Bruno Lima} {\,} is an Assistant Professor in Software Engineering at UMaia and FEUP. He is also a researcher at LIACC. Contact him at brunolima@fe.up.pt.
\end{IEEEbiography}

\vspace*{-10pt}

\begin{IEEEbiography}{João P. Faria} {\,} is a full professor in Software Engineering at FEUP and a senior researcher at INESC TEC. He is an IEEE member. Contact him at jpf@fe.up.pt.
\end{IEEEbiography}

\end{document}